\documentclass[fleqn,usenatbib]{mnras}

\usepackage{newtxtext,newtxmath}
\usepackage[T1]{fontenc}

\DeclareRobustCommand{\VAN}[3]{#2}
\let\VANthebibliography\thebibliography
\def\thebibliography{\DeclareRobustCommand{\VAN}[3]{##3}\VANthebibliography}

\usepackage{graphicx}	% Including figure files
\usepackage{amsmath}	% Advanced maths commands

\title[Galactic NSWDs as GW sources]{Constraining Mass Transfer Models with Galactic Neutron Star$-$White Dwarf Binaries as Gravitational Wave Sources}

\author[Jian-Guo He et al.]{
Jian-Guo He$^{1,2}$,
Yong Shao$^{1,2}$\thanks{E-mail: shaoyong@nju.edu.cn},
Xiao-Jie Xu$^{1,2}$,
Xiang-Dong Li$^{1,2}$
\\
$^{1}$Department of Astronomy, Nanjing University, Nanjing 210023, People's Republic of China\\
$^{2}$Key Laboratory of Modern Astronomy and Astrophysics, Nanjing University, Ministry of Education, Nanjing 210023, People's Republic of China
}

\date{Accepted XXX. Received YYY; in original form ZZZ}

\pubyear{2023}

\begin{document}
\label{firstpage}
\pagerange{\pageref{firstpage}--\pageref{lastpage}}
\maketitle

% Abstract of the paper
\begin{abstract}
Neutron star$-$white dwarf (NSWD) binaries are  one of the most abundant sources of gravitational waves (GW) in the Milky Way. These GW sources are the evolutionary products of primordial binaries that experienced many processes of binary interaction. We employ a binary population synthesis method to investigate the properties of Galactic NSWD binaries detectable by the Laser Interferometer Space Antenna (LISA). In this paper, only the NSWD systems with a COWD or ONeWD component are included. We consider various models related to mass transfer efficiencies during primordial binary evolution, supernova explosion mechanisms at NS formation, common envelope ejection efficiencies, and critical WD masses that determining the stability of mass transfer between WDs and NSs. Based on our calculations, we estimate that tens to hundreds of LISA NSWD binaries exist in the Milky Way. We find that the detection of LISA NSWD binaries is able to provide profound insights into mass transfer efficiencies during the evolution of primordial binaries and critical WD masses during mass transfer from a WD to an NS. 
\end{abstract}

% Select between one and six entries from the list of approved keywords.
% Don't make up new ones.

%%%%%%%%%%%%%%%%%%%%%%%%%%%%%%%%%%%%%%%%%%%%%%%%%%

%%%%%%%%%%%%%%%%% BODY OF PAPER %%%%%%%%%%%%%%%%%%

\section{Introduction}  \label{sec:introduction}

In recent years, ground-based gravitational wave (GW) detectors such as LIGO and Virgo have identified nearly one hundred mergers of double compact objects \citep{Abbott2023}, primarily composed of black holes (BHs), since the groundbreaking detection of GW150914 \citep{Abbott2016}. Among these mergers, two (GW170817 and GW190425) originate from double neutron star (NS) systems. To date, no event involving the merger of an NS and a white dwarf (WD) has been confirmed. 
The detection of GW signals from NSWD binaries is able to help resolve some astrophysical problems, including the stability of mass transfer between WDs and NSs \citep[e.g.,][]{Verbunt1988,Bobrick2017}, the equation of state of NS matter \citep{Tauris2018}, the possible origins of ultra-compact X-ray binaries \citep[UCXBs,][]{Haaften2013,Wang2021}, repeating fast radio bursts \citep{GuWM2016}, faint type Iax supernovae \citep{Bobrick2022}, and peculiar gamma-ray bursts \citep[e.g.,][]{YangJ2022,Kaltenborn2023}. Future space-based GW detectors such as LISA \citep{AS2017} and TianQin \citep{Luo2016} are promising to detect these GW signals in the mHz band. 

It is expected that the Milky Way hosts hundreds of NSWD systems observable by LISA \citep{AS2023}. 
An early work by \citet{Nelemans2001b} indicated that LISA may detect several hundreds of Galactic NSWD binaries. Based on the observations of close binary radio millisecond pulsars, \citet{Tauris2018} inferred that at least a hundred NSWD binaries with helium WDs (HeWDs) could be detected as LISA sources in the Milky Way. In addition, \citet{ChenWC2020} estimated the existence of approximately 60$-$80 interacting LISA NS$-$HeWD systems in the Galactic field. More recently, \citet{Korol2023} suggested that around 80$-$350 detached Galactic NSWDs are detectable by LISA over its 4-year duration, depending on models related to the kick velocities of natal NSs and the treatments of common envelope (CE) evolution. Outside the Milky Way, LISA is likely to  detect about $ 1-6$ NSWD systems in M31 for a 10-year mission  \citep{HeJG2023}. In galaxies beyond the local group, GW signals from merging NSWD binaries are challenging to observe due to their large distances and limited chirp masses \citep{AS2023}, unless the operation of future next-generation detectors such as DO-OPT and DEC \citep{Kang2024}.

Based on the canonical  channels with isolated binary evolution, close NSWD systems are expected to be the descendants of low-mass X-ray binaries that experienced stable Roche lobe overflow (RLOF) or intermediate-mass X-ray binaries that experienced CE evolution \citep{Tauris2023}. The former channel always results in the formation of HeWDs, while the latter tends to produce more massive WDs, i.e., carbon-oxygen WDs (COWDs) or oxygen-neon WDs (ONeWDs). In some cases, it is possible that mass transfer during the progenitor evolution of NSWD binaries can effect a reversal of the end states of the two components, resulting in a WD that forms before an NS \citep[e.g.,][]{Nelemans2001b}. Observations of detached NSWD systems with orbital periods of $< 0.4$ days and HeWD masses of $< 0.2M_{\odot}$ pose a challenge to the RLOF channel since the formation of these binaries is very sensitive to various factors such as initial NS masses, NS's accretion efficiencies, and magnetic braking mechanisms \citep{Istrate2014,ChenWC2020}. As the orbits of these detached NSWD binaries significantly shrink due to GW radiation, they are likely to become detectable GW sources. Subsequently, these systems evolve to be UCXBs when the WD companion fills its Roche lobe and transfers material to the NS. It has been shown that the duration approximately one million years before and after the onset of UCXBs represents a GW detection window \citep{Tauris2018}, indicating an overlap in the evolutionary pathways of NSWD binaries as LISA sources and UCXBs. In recent investigations on Galactic NSWD binaries as GW sources \citep{Tauris2018, ChenWC2020, Korol2023}, UCXBs with HeWDs and detached systems with  HeWDs/COWDs/ONeWDs have been considered. It is possible that UCXBs with COWDs/ONeWDs also contribute to the population of GW sources if a massive WD can stably transfer matter  to an NS.

The stability of mass transfer between WDs and NSs has been extensively studied but remains uncertain. The traditional jet-only model \citep{Verbunt1988} suggests a critical WD mass of approximately $0.5 M_{\odot}$ for stable mass transfer. Later, the isotropic re-emission mass-transfer assumption gives a  limit of around $0.4 M_{\odot}$ for WD masses \citep{Yungelson2002, Haaften2012}, taking into account the inability to eject a sufficient amount of transferred matter from the system, further reducing the stability of mass transfer. By assuming that the NSWD systems with WDs of masses $> 0.38 M_{\odot}$ lead to unstable mass transfer and merge, \citet{Haaften2013}  pointed out that most UCXBs consist of an HeWD component. Based on hydrodynamic simulations, \citet{Bobrick2017} revealed a lower critical WD mass of $0.2 M_{\odot}$ when considering the effect of disc winds that developed at highly super-Eddington mass-transfer rates. In this case, only NSWD systems containing an HeWD can evolve into UCXBs. Using the same method, \citet{Church2017} obtained a critical value of approximately $0.15 - 0.25 M_{\odot}$ for WD masses, depending on the assumptions of initial input parameters and specific physical models. However, in contrast to these results, \citet{ChenHL2022} suggested that all NS$-$HeWD binaries with WD masses of $0.17-0.45M_\odot$ are expected to undergo stable mass transfer when using the detailed stellar evolution code MESA, which takes into account the realistic structure of WDs. \citet{ChenHL2022} also demonstrated that the stability of mass transfer from an HeWD to an NS is independent of NS's mass and its mass-accretion efficiency that characterizes the fraction of transferred matter accreted by the NS. In a different approach, \citet{Yu2021} developed a code to investigate the orbital evolution of mass-transferring NSWD binaries and showed that the majority of these systems experience stable mass transfer. They found that the maximum WD mass can reach approximately $1.25-1.28 M_{\odot}$ for stable mass transfer, beyond which NSWDs directly merge when mass transfer begins.

In this paper, we provide a comprehensive study on the characteristic distribution of Galactic NSWD binaries as GW sources. Using a binary population synthesis \citep[BPS, see a review by][]{Han2020} method, we consider the effect from various aspects such as the treatments of mass transfer between binary components, the efficiencies of CE ejection, and the mechanisms of supernova explosion. % to explore their formation channels, total numbers, component masses, eccentricity, and Galactic scale height. 
The structure of this paper is organized as follows. In Section \ref{sec:method}, we present the methodology employed, which includes the adoption of different models. Subsequently, in Section \ref{sec:result}, we present the results derived from our calculations. In Section \ref{sec:discussion}, we make some discussions based on these results. Finally, we conclude in Section \ref{sec:conclusion}.

\section{Method}   \label{sec:method}

We employ the BSE code developed by \citet{ Hurley2002} to investigate the population of Galactic GW sources of NSWD binaries across various models. These models involve different supernova recipes, mass-transfer efficiencies and its stability, as well as physical parameters related to CE evolution. Additionally, we account for the star formation history of the Milky Way and perform the integration of the spatial motion of NSWDs under the influence of Galactic gravitational potential. Taking into account the spatial locations of Galactic NSWD systems, we calculate the signal-to-noise ratio (S/N) for every GW binary in the LISA band and obtain the population of detectable sources accordingly.

\subsection{Supernova Mechanisms}

Regarding the types and the masses of stellar remnants following core-collapse supernova (CCSN) explosions, we utilize the rapid mechanism \citep{Fryer2012}, which correlates the remnant masses with the masses of the CO cores prior to explosions. This mechanism can account for the existence of the mass gap between NSs and BHs \citep{Shao2022}. Under this mechanism, we adopt a Maxwell distribution with a standard deviation of $\sigma=265 \mathrm{~km} \mathrm{~s}^{-1}$ \citep{Hobbs2005} for the kick velocities of natal NSs. For NSs formed through electron capture supernovae (ECSN) or accretion-induced collapse (AIC), we use smaller kick velocities with $\sigma=30 \mathrm{~km} \mathrm{~s}^{-1}$ \citep{Vigna2018, Shao2018}.

In addition, we consider an alternative supernova explosion mechanism with the stochastic recipe proposed by \citet{Mandel2020}. Unlike the rapid mechanism, this recipe introduces randomness in compact remnant masses and kick velocities. Both the rapid and the stochastic recipes have been incorporated into the BSE code \citep{Shao2021}.

\subsection{Mass Transfer}

In a binary, the stability of mass transfer is determined by considering the adiabatic hydrostatic response of the donor star to mass loss, denoted as $\zeta_{\mathrm{ad}}$, 
\begin{equation}
\zeta_{\mathrm{ad}}=\left.\frac{\partial \ln R_2}{\partial \ln M_2}\right|_{\mathrm{ad}},
\end{equation}
as well as the response of the Roche lobe to mass loss, denoted as $\zeta_{\mathrm{RL}}$,
\begin{equation}
\zeta_{\mathrm{RL}}=\left.\frac{\partial \ln R_{\mathrm{L,2}}}{\partial \ln M_2}\right|_{\mathrm{bin}},
\end{equation} 
where $R_2$ and $R_{\mathrm{L,2}}$ are the radii of the donor star and its Roche lobe, respectively, and $M_2$ is the mass of the donor star \citep[see e.g.,][]{Soberman1997}. When $\zeta_{\mathrm{RL}}<\zeta_{\mathrm{ad}}$, mass transfer proceeds in a stable manner. Otherwise, dynamically unstable mass transfer occurs and CE evolution is triggered.

\subsubsection{Efficiency of mass transfer}

During the evolution of the primordial binaries initially containing two zero-age main-sequence stars, mass-transfer efficiency ($\eta_{\mathrm{MT}}$) characterizes the fraction of the transferred matter that is accreted by the secondary star, which 
plays a crucial role in determining whether the binaries undergo stable  mass transfer or evolve into a contact/CE phase \citep[e.g.,][]{deMink2007}.
It has been demonstrated that a lower mass-transfer efficiency tends to prevent significant expansion of the secondary star due to accretion, allowing a larger parameter space of primordial binaries for stable mass transfer. Based on the work of \citet{Shao2014}, we employ three mass accretion models to deal with the process of mass transfer during primordial binary evolution. By default, we utilize the rotation-dependent mass accretion model, referred to as MA1. In this model, the mass accretion rate of the secondary star is assumed to be dependent on its rotational velocity, so the mass-transfer process could be highly non-conservative with $\eta_{\mathrm{MT}}\lesssim 20\%$. Alternatively, we consider two other models: half mass accretion and thermal equilibrium-limited mass accretion, referred to as MA2 and MA3, respectively, corresponding to $\eta_{\mathrm{MT}}=50\%$ and $\eta_{\mathrm{MT}} \sim 100\%$. Each accretion model is associated with a specific criterion to decide the stability of mass transfer between binary components \citep{Shao2014}. Previous investigations have shown that the rotation-dependent model can better match the observations of Galactic OBe-star binaries with a BH or a Wolf-Rayet star companion, while the observations of Galactic Be-star binaries with an NS or a subdwarf companion seem to favor the models with  $\eta_{\mathrm{MT}} \gtrsim 0.5$ \citep[][and references therein]{Shao2022}.

When the accretor is an NS, we assume an accretion efficiency of 0.5 \citep{ChenWC2020}. In addition, the accretion rate of the NS is constrained by the Eddington limit. In our calculations, we adopt the isotropic re-emission mechanism for non-conservative mass transfer. It is assumed that the material lost from a binary system carries away the specific angular momentum of the accretor.

\subsubsection{Mass transfer from a nondegenerate star to an NS}

In the case of non-conservative mass transfer with an isotropic re-emission way, previous works \citep[e.g.,][]{Soberman1997} have suggested a positive correlation between $\zeta_{\mathrm{RL}}$ and $q = M_{\rm d}/M_{\rm NS}$ (mass ratio of the donor to the NS). As a consequence, there are critical mass ratios used to determine mass-transfer stability. \citet{Tauris2000} pointed out that the NS binaries with $q \gtrsim 3-3.5$ always evolve into CE phases while the systems with $q \lesssim 1.5-2$ are expected to undergo stable mass transfer \citep[see also][]{Shao2012,Misra2020}. Based on detailed evolutionary simulations of the BH binaries with nondegenerate donors, \citet{Shao2021} obtained easy-to-use criteria for mass-transfer stability. In this paper, we apply these criteria to the binaries with an NS accretor. It is assumed that mass transfer is always stable if $q<2$ or always unstable if $q>2.1+0.8M_{\rm NS}$. For the systems with mass ratios between them, mass transfer stability is dependent on the properties of the donor stars.

For the binaries with a naked He star and an NS, we adopt the criteria calculated by \citet{Tauris2015} to deal with mass-transfer stability. According to their work, CE phases are expected to occur when the masses of the He stars exceed $2.7 M_{\odot}$ and the orbital periods are below $0.06$ days. So we assume that only the systems with He-star masses of $>2.7 M_{\odot}$ and orbital periods of $<0.06$ days evolve into CE phases.

\subsubsection{Mass transfer from a WD to an NS}

Considering that UCXBs with a WD donor and an NS accretor can last $\sim 1\,\rm Myr$ as GW sources, it is speculated that the stability of mass transfer between WDs and NSs significantly impacts the population properties of LISA NSWD binaries. Until now, however, the stability of mass transfer from a WD  to an NS still remains uncertain. By default, we  adopt a threshold of $0.2 M_{\odot}$ for WD masses \citep{Bobrick2017} to investigate the properties of detached NSWD systems observable by LISA. This threshold does not allow the NS binaries with COWDs/ONeWDs to evolve into long-standing UCXBs. Also, we vary this threshold mass $M_{\mathrm{WD, crit}}$ from $0.4 M_{\odot}$ \citep{Haaften2012} to $1.25 M_{\odot}$ \citep{Yu2021} and explore its influence on the number of interacting NSWD systems (UCXBs) in the Milky Way.

For other types of binary systems not mentioned above, we use the default criteria given by \citet{Hurley2002} to deal with the stability of mass transfer.

\subsubsection{CE Evolution}

When CE evolution is triggered, we employ the energy conservation prescription to determine the outcome of CE phases. The related formulae can be found in \citet{Hurley2002} and \citet{Shao2014}. In our study, we utilize the binding energy parameter $\lambda$ fitted by \citet{XuXJ2010}. By default, we assume the efficiency of CE ejection to be unity, i.e., $\alpha_{\mathrm{CE}}=1.0$. This parameter determines the proportion of orbital energy lost that used to eject donor's envelope. To assess the impact of $\alpha_{\mathrm{CE}}$ on the population of LISA NSWD systems, we consider two additional efficiencies, by choosing $\alpha_{\mathrm{CE}}=0.3$ as inferred from the parameter distribution of Galactic WDWD systems \citep{Scherbak2023} and $\alpha_{\mathrm{CE}}=3.0$ as required by the formation of the post-CE system IK Pegasi \citep{Davis2010}.

\subsection{Primordial Binaries}

In our study, we simulate the evolution of approximately $10^6$ primordial binaries for each model. The initial parameters of primordial binaries are set as follows. The primary masses range from $5$ to $100 M_{\odot}$, and the secondary masses range from $0.5$ to $100 M_{\odot}$. All primordial binaries are assumed to have circular orbits, with separations varying from $3$ to $10000 R_{\odot}$. The binary fraction among all stars is assumed to be unity. We follow the method of \citet{Shao2021} to calculate the Galactic population of LISA NSWD systems that evolved from primordial binaries.

\subsection{Star Formation History and Orbital Integration}

For the Milky Way, we assume a constant star formation rate of $3 M_{\odot} \mathrm{~yr}^{-1}$ and a constant metallicity of $Z_{\odot}=0.02$ throughout its entire lifespan of 10 $\mathrm{Gyr}$. 

To account for the spatial motions of NSWD binaries, we utilize the galpy package \citep[with the MWPotential2014 model,][]{Bovy2015} to numerically integrate their tracks from the formation of NS until either the binaries merge or the evolutionary time exceeds 10 $\mathrm{Gyr}$. 

Regarding the initial locations of primordial binaries in the Milky Way, the star number density distribution can be described as a function of radial distance from the Galactic center $r$ and vertical distance from the Galactic plane $z$, using the equation proposed by \citet{Bahcall1980} as
\begin{equation}
\rho(r, z) = \rho_{\odot} \exp \left[-\frac{r-r_{\odot}}{h_r}-\frac{z}{h_z}\right],
\end{equation}
where $r_{\odot}=8.5 \mathrm{~kpc}$ represents the radial distance of the Sun from the Galactic center, and $\rho_{\odot}$ denotes the star number density at the location of the Sun. Here, $h_r=3.5 \mathrm{~kpc}$ and $h_z=0.25 \mathrm{~kpc}$ represent the scale length parallel and perpendicular to the Galactic plane. In the Galactic reference frame, the velocities of the mass centers of pre-SN systems are assumed to be consistent with rotation curves, where the circular velocity of the Sun is 220 $\mathrm{km} \mathrm{~s}^{-1}$. 

Following the approach of \citet{Hurley2002}, the velocities of the new mass centers of post-SN binaries are subject to changes caused by mass losses and NS kicks during supernova explosions.

\subsection{Signal-to-noise Ratio (S/N) of GW}

We utilize the python package LEGWORK \citep{Wagg2022} to calculate the S/N of Galactic NSWD binaries and identify those with an S/N greater than 5 as detectable LISA sources. Among optional spacecraft parameters, we select the robson19 model \citep{Robson2019} for the sub-mHz confusion noise contributed by unresolved WDWD binaries in the Galaxy \citep{Cornish2017, Babak2021}. By default, the observation time is set to 4 years, which is the standard duration used in LISA mission simulations.

\section{Result} \label{sec:result}

It is challenging for rapid population synthesis to model the formation of close NSWD systems with low-mass HeWDs that involving the RLOF channel and requiring severe fine-tuning of input parameters \citep[see e.g.,][]{Istrate2014,ChenWC2020,Deng2021}. Consequently, the NS$-$HeWD systems as GW sources are absent in our results\footnote{Despite the absence of NS$-$HeWD systems in our results, we can evaluate their impact on LISA detection. Previous studies have demonstrated that LISA may detect more than 100 NS$-$HeWD binaries in the Milky Way \citep{Tauris2018,ChenWC2020}. Magnetic braking mechanisms are thought to play a crucial role in forming close NS$-$HeWD systems and alleviating the fine-tuning problem \citep{Deng2021,ChenHL2021}. For LISA NSWD sources, one may differentiate HeWDs from COWDs/ONeWDs with measured chirp masses in detached systems or observed spectral lines in interacting systems. Possible detection of LISA NS$-$HeWD systems can be used to constrain the mechanisms of magnetic braking.}. And, we consider LISA NSWD systems with NSs originating from CCSN and ECSN, while those with NSs originating from AIC are disregarded here and discussed in Section \ref{sec:AIC Mechanism}.

Our calculations reveal that the total number of Galactic NSWD systems in the Milky Way is about $2\times10^6$, which is consistent with the estimation of \citet{Nelemans2001b}. Expected numbers of detectable NSWD systems by LISA vary across different models, as presented in Table \ref{tab:total number}. In this table, we separately list the numbers of detached and interacting LISA sources, as well as the merger rates of NSWD systems.

In Section \ref{sec:channel}, we discuss the evolutionary pathways and initial binary parameter spaces to form LISA NSWD binaries. Subsequently, in Section \ref{sec:detached}, we evaluate the influence of different models related to the options of $\eta_{\mathrm{MT}}$, $\alpha_{\mathrm{CE}}$, and supernova recipes. We analyze the distributions of various parameters of LISA NSWD binaries including their orbital  parameters, component masses, and Galactic locations. In Section \ref{sec:interacting}, we investigate the impact of $M_{\mathrm{WD, crit}}$ on interacting LISA NSWDs. In Section \ref{sec:merger}, we estimate the merger rates of NSWD binaries in the Milky Way and the local Universe.

\begin{table*}
	\centering
	\caption{Expected numbers and merger rates ($R_{\mathrm{merger}}$) of LISA NSWD binaries in the Milky Way. Our models include different treatments of mass-transfer efficiencies during primordial binary evolution ($\eta_{\mathrm{MT}}$), critical WD masses for the stability of mass transfer between WDs and NSs ($M_{\mathrm{WD,cri}}$), CE ejection efficiencies ($\alpha_{\mathrm{CE}}$) and supernova recipes. MA1, MA2, and MA3 represent rotation-dependent mass accretion, half mass accretion ($\eta_{\mathrm{MT}}=50\%$), and near-conservative mass accretion ($\eta_{\mathrm{MT}}\sim 100\%$), respectively. The superscripts D and I refer to detached and interacting LISA NSWD binaries, respectively.  $\mathcal{R}_{\mathrm{merger}}$ denotes estimated merger rate densities of NSWD systems in the local Universe.}
	\label{tab:total number}
	\begin{tabular}{cccccccccc}
%    \hline
%    \multicolumn{4}{c}{Models} & \multicolumn {5}{c}{Number} \\
    \hline
     $\eta_{\mathrm{MT}}$ &  $M_{\mathrm{WD,cri}}$ &  $\alpha_{\mathrm{CE}}$ & SN recipes  & $N^{\mathrm{D}}$ & $N^{\mathrm{I}}$ & $N^{\mathrm{D}}$ \textbf{($N^{\mathrm{I}}$)} & $N^{\mathrm{D}}$ \textbf{($N^{\mathrm{I}}$)} & $R_{\mathrm{merger}}$ & $\mathcal{R}_{\mathrm{merger}}$ \\
      & $(M_{\odot})$ &  &  &  &  &  NS forms first & WD forms first & $(\mathrm{Myr}^{-1})$ & $(\mathrm{Gpc}^{-3} \mathrm{yr}^{-1})$\\
    \hline
    MA1 & 0.2 & 0.3 & rapid & 45 & $-$ & 34 & 11 & 17.4 & 174\\
    MA1 & 0.2 & 1.0 & rapid & 105 & $-$ & 79 & 26 & 43.3 & 433\\
    MA1 & 0.2 & 1.0 & stochastic & 162 & $-$ & 120 & 42 & 66.9 & 669\\
    MA1 & 0.2 & 3.0 & rapid & 197 & $-$ & 149 & 48 & 83.3 & 833\\
    MA2 & 0.2 & 0.3 & rapid & 17 & $-$ & 14 & 3 & 7.2 & 72\\
    MA2 & 0.2 & 1.0 & rapid & 78 & $-$ & 55 & 23 & 31.2 & 312\\
    MA2 & 0.2 & 1.0 & stochastic & 103 & $-$ & 72 & 31 & 42.3 & 423\\
    MA2 & 0.2 & 3.0 & rapid & 153 & $-$ & 86 & 67 & 64.3 & 643\\
    MA3 & 0.2 & 0.3 & rapid & 49 & $-$ & 11 & 38 & 18.1 & 181\\
    MA3 & 0.2 & 1.0 & rapid & 145 & $-$ & 47 & 98 & 55.9 & 559\\
    MA3 & 0.2 & 1.0 & stochastic & 211 & $-$ & 63 & 148 & 82.6 & 826\\
    MA3 & 0.2 & 3.0 & rapid & 234 & $-$ & 64 & 170 & 88.9 & 889\\  
    MA1 & 1.25 & 1.0 & rapid & 105 & 194 & 79 (174) & 26 (20) & 5.1 & 51\\
    MA2 & 1.25 & 1.0 & rapid & 78 & 116 & 55 (100) & 23 (16) & 8.7 & 87\\
    MA3 & 1.25 & 1.0 & rapid & 145 & 120 & 47 (99) & 98 (21) & 31.8 & 318\\
    \hline
  \end{tabular}
\end{table*}

\subsection{Formation Scenarios and Binary Parameter Spaces}  \label{sec:channel}

NSWD systems can be categorized based on the formation order of their binary components, specifically whether the NS or the WD forms first. Therefore, we classify the formation scenarios for LISA NSWD sources as follows:

\textit{Scenario 1:} NS forms first.

\textit{Scenario 2:} WD forms first. 

Evolved from primordial binaries, Scenario 1 means that the primary stars firstly evolve into NSs and then the secondary stars become WDs. There is a common feature that the NSWD binaries formed via Scenario 1 have circular orbits. The masses of the primary stars and the secondary stars fall within the ranges of $6-20 M_{\odot}$ and $2-10 M_{\odot}$, respectively. The  orbital periods of the primordial binaries cover two separated ranges, i.e., a few days to tens of days and hundreds to thousands of days. Generally, the primordial binaries with narrow orbits undergo stable mass transfer during the first mass-transfer stage, while those with wide orbits are expected to experience CE evolution since the primary stars have developed a deep convective envelope prior to mass transfer.

In Scenario 2, the primary stars typically have initial masses of $\gtrsim 5 M_{\odot}$ and the secondary stars have similar masses. As the progenitors of LISA NSWD sources, the corresponding primordial binaries have relatively narrow orbits with periods of a few days to tens of days. During the evolution, mass transfer occurs through stable RLOF. The primary stars firstly  turn into WDs. Since the secondary stars have accreted sufficient matter during previous mass-transfer processes, they can evolve into He stars with masses of $\gtrsim 2M_{\odot}$ after their hydrogen envelopes stripped via CE phases,  and eventually become NSs. Additionally, it is possible for Scenario 2 that a small fraction of the primordial binaries with long orbital periods evolve to be double He-star systems before WD and NS formation \citep[see also the Pathway 3 described by][]{Toonen2018}. In contrast to Scenario 1, NSWD binaries where WDs form first tend to exhibit large orbital  eccentricities due to the lack of an efficient mechanism of orbital circularization after NS formation.

\subsection{Detached NSWD Binaries}  \label{sec:detached}

\subsubsection{The impact of mass-transfer efficiency}

\begin{figure*}
	\includegraphics[width=18cm]{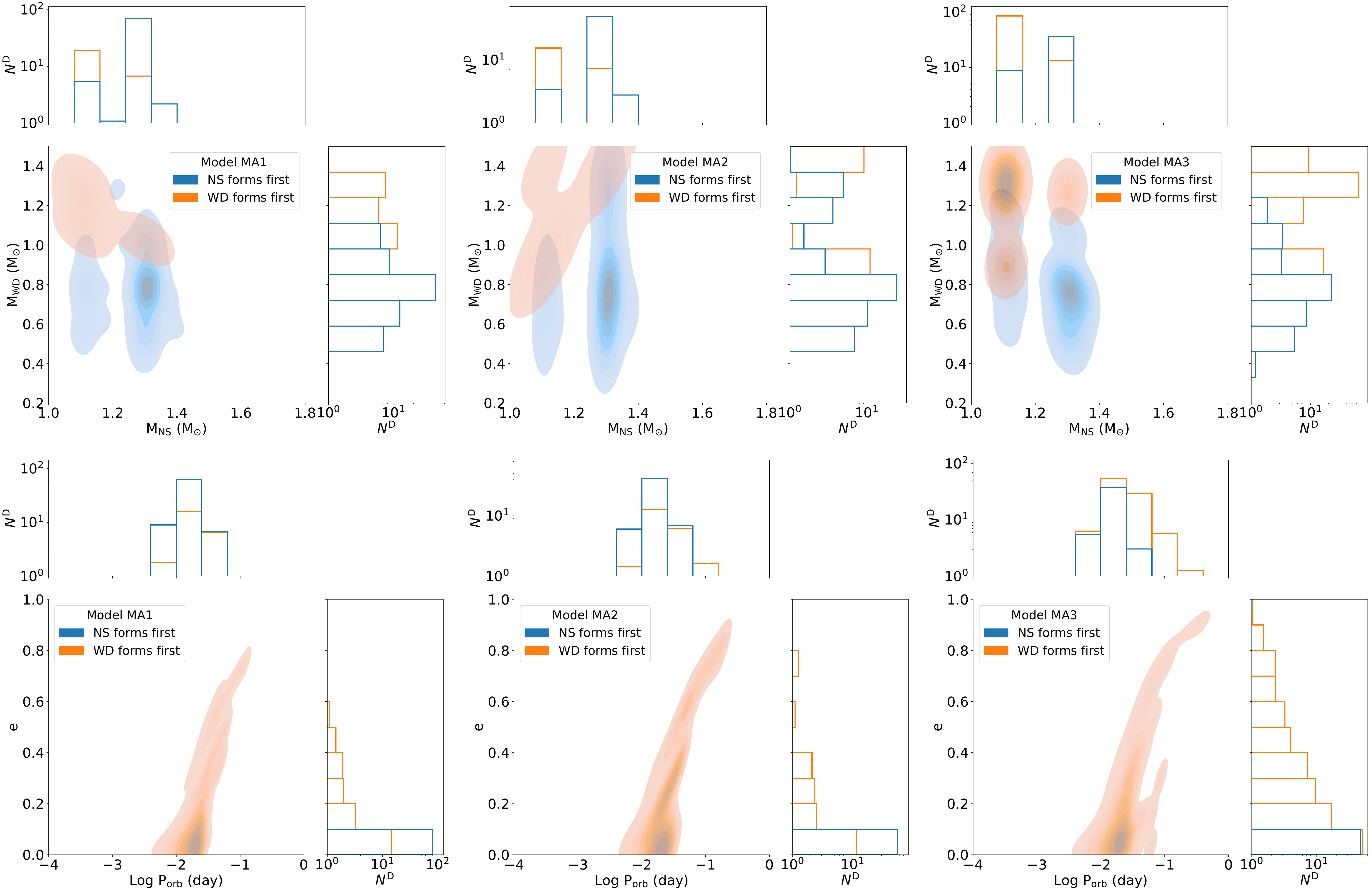}
    \caption{Calculated number distributions of detached LISA NSWD systems in the Milky Way, as a function of NS mass, WD mass, orbital period, and eccentricity. The left, middle, and right panels correspond to the models MA1, MA2 and MA3, respectively. Here, we adopt $\alpha_{\mathrm{CE}} = 1$ and the rapid mechanism of supernova explosions. In each panel, the blue contours represent systems where NS forms first, while the orange contours represent systems where WD forms first. Notably, all NSWD binaries where NS forms first exhibit circular orbits, the corresponding blue contours do not appear to show these systems in the plane of orbital period versus eccentricity.} 
    \label{fig:detached}
\end{figure*}

Fig. \ref{fig:detached} presents calculated number distributions of Galactic LISA sources of detached NSWD binaries in the planes of NS mass versus WD mass (upper panels) and orbital period versus eccentricity (lower panels). The left, middle, and right panels correspond to the mass accretion models MA1, MA2, and MA3, respectively. In this analysis, we adopt the default rapid supernova explosion mechanism and set $\alpha_{\mathrm{CE}} = 1$. The value of $M_{\mathrm{WD, crit}}$ is fixed at $0.2 M_{\odot}$, resulting in all LISA NSWD systems being detached. The reason is that when NSs form first, WD masses are typically above $0.4 M_{\odot}$, whereas when WDs form first,  WD masses are generally larger than $0.8 M_{\odot}$. This also indicates the absence of HeWDs in all cases. Since the critical mass ratio of nondegenerate donors to NSs for avoiding CE evolution is $\sim 2$ \citep[see e.g.,][]{Misra2020}, LISA NSWD binaries formed via Scenario 1, as the evolutionary products of intermediate-mass X-ray binaries, are expected to host COWDs or ONeWDs. On the other hand, NSWD systems formed via Scenario 2 require the masses of both components of the primordial binaries to exceed $5 M_{\odot}$, leading to produce massive WDs. The masses of NSs are distributed with two  peaks at $\sim 1.1 M_{\odot}$ and $\sim 1.3 M_{\odot}$, corresponding to the NSs formed from CCSN and ECSN, respectively. For CCSN NSs, we adopt the rapid mechanism \citep{Fryer2012} which predicts that stars with masses of $\sim 8-12 M_{\odot}$ finally collapse into  $\sim 1.1 M_{\odot}$ NSs. For ECSN NSs, we simply assume that they are born with mass of $ 1.3M_{\odot}$ \citep[see also][]{HeJG2023}.

It is noteworthy that distinguishing the evolutionary origins of detached NSWD systems from the RLOF and the CE channels is relatively straightforward. The  RLOF channel is expected to produce LISA NS$-$HeWD binaries with chirp masses of $\lesssim 0.44 M_{\odot}$, corresponding to the systems containing a $\lesssim 2 M_{\odot}$ NS and a $\sim 0.167 M_{\odot}$ WD \citep{Tauris2018}. For the CE channel, our calculations indicate a minimum chirp mass of $\sim 0.56 M_{\odot}$ for LISA NSWD systems which contain a $\gtrsim 1.1 M_{\odot}$ NS and a $\gtrsim 0.4 M_{\odot}$ WD. Consequently, we can confidently discern the evolutionary channels of LISA NSWD systems. Next, we focus on the systems with COWD/ONeWD components and perform quantitative analyses of binary parameters under different models.

Among three mass accretion models, we see that MA3, which corresponds to nearly conservative mass transfer, yields the highest number of LISA NSWD systems where WDs form first. Additionally, the majority of these systems are expected to host an ONeWD. As mass-transfer efficiency increases, the component masses of primordial binaries in Scenario 2 shift towards lower values, resulting in the formation of more systems where WDs form first due to initial mass function. Specifically, the model MA3 predicts the existence of approximately 50 detached NSWD systems with eccentricities exceeding 0.1. In this case, the fraction of these eccentric binaries among all detached LISA systems with COWDs/ONeWDs (the number is 145 from Table \ref{tab:total number}) can reach as high as $\sim 0.34$.
And, we estimate that $\sim 8$ systems are likely to have relatively wide orbits with periods ranging from 0.1 to 1 day. In contrast, the  models MA1 and MA2 predict that almost no LISA NSWD binaries have large eccentricities of $> 0.5$, and only around 10 systems are expected to have eccentricities exceeding 0.1, corresponding to their fraction of $\sim 0.1-0.14$ among all $78-105$ LISA binaries (see Table \ref{tab:total number}). Based on the distributions of orbital parameters combined with the numbers of detached NSWD systems detectable by LISA, it becomes feasible to provide constraints on possible mass accretion models from future GW observations. 

\begin{figure*}
	\includegraphics[width=15cm]{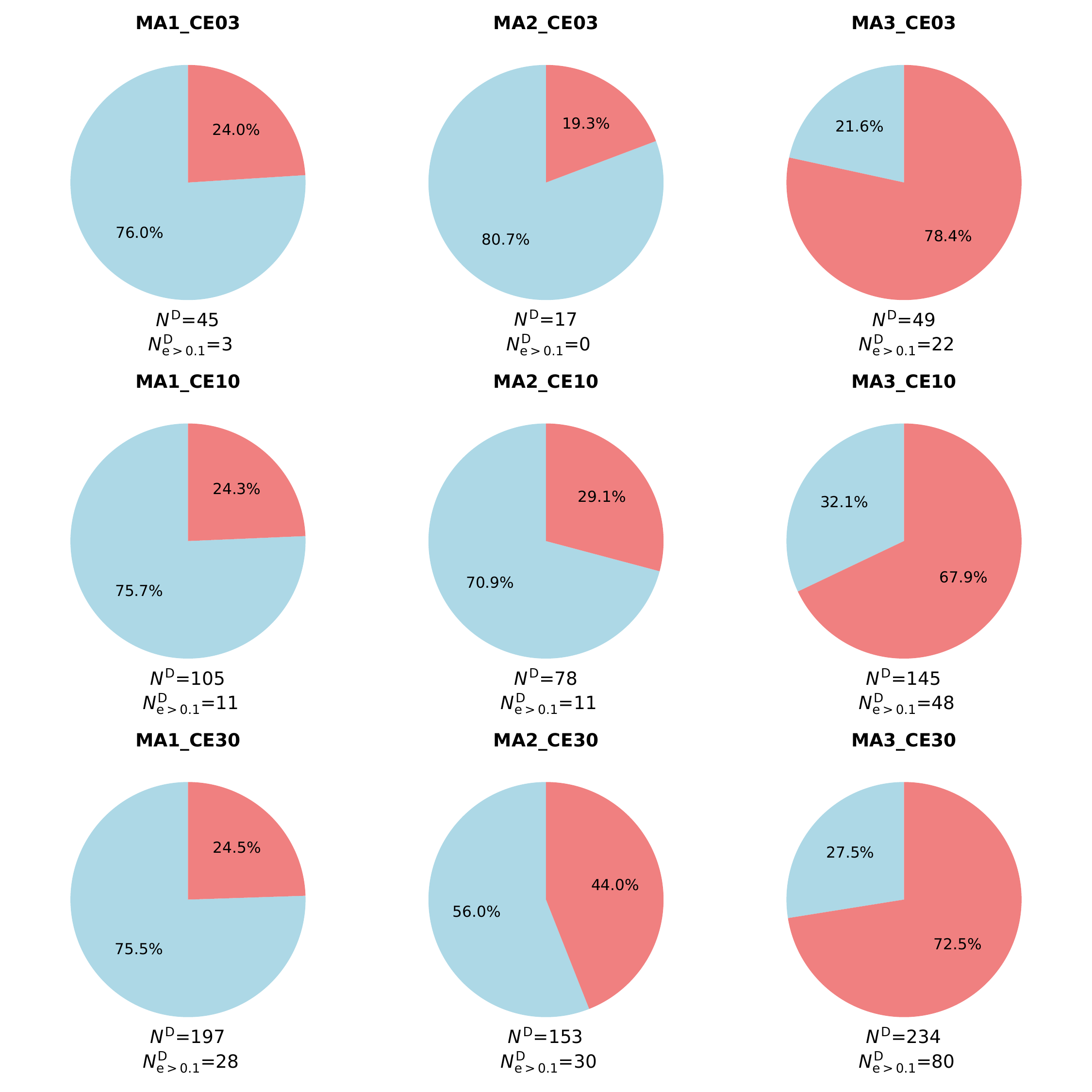}
    \caption{Pie charts illustrating the relative fractions of detached LISA NSWD systems where NSs (blue) or WDs (red) form first. The left, middle, and right panels represent the models MA1, MA2 and MA3, respectively. The panels from top to bottom correspond to  $\alpha_{\mathrm{CE}}=0.3$, 1.0, and 3.0, respectively. Below each pie chart, we also give the corresponding numbers for all detached binaries and the systems with eccentricities larger than 0.1.} 
    \label{fig:pie}
\end{figure*}

\subsubsection{The impact of $\alpha_{\mathrm{CE}}$}

Fig. \ref{fig:pie} shows the relative fractions of detached LISA NSWD binaries formed via Scenario 1 or 2. The left, middle, and right panels represent the models MA1, MA2 and MA3, respectively. The panels from top to bottom correspond to $\alpha_{\mathrm{CE}}=0.3,$ 1.0, and 3.0, respectively. In each panel, we also give the  numbers of all detached LISA binaries from our calculations and the systems with eccentricities larger than 0.1.

It is obvious that mass-transfer  efficiencies during primordial binary evolution are the dominant factor influencing the relative fractions of systems where NSs or WDs form first. However, we note that the influence of $\alpha_{\mathrm{CE}}$ cannot be disregarded. On the one hand, a lower value of $\alpha_{\mathrm{CE}}$ make it more challenging to eject CE, resulting in a significant reduction in the number of systems. Expected  numbers of LISA NSWD binaries ($N^{\rm D}$) decrease from 153$-$234, to 78$-$145, and to 17$-$49 when decreasing $\alpha_{\mathrm{CE}}$ from 3.0, to 1.0, and to 0.3, respectively. On the other hand, the numbers of the systems with eccentricities above 0.1 ($N^{\rm D}_{\rm e>0.1}$) are sensitive to the options of $\alpha_{\mathrm{CE}}$ in the models MA1 and MA2. Overall, the ratios of $N^{\rm D}_{\rm e>0.1}$ to $N^{\rm D} $ are always $ \lesssim 0.2$ in these two models. While for all our adopted $\alpha_{\mathrm{CE}}$, the model MA3 predicts that $\sim 0.3-0.4$ of detectable NSWD binaries have eccentricities larger than 0.1. This is because the primordial binaries in Scenario 2 have the orbital periods of 8$-$27 days for the model MA3, while they have the orbital periods of  2$-$20 days for the models MA1 and MA2. The latter is more susceptible to the influence of $\alpha_{\mathrm{CE}}$ when CE evolution occurs in the subsequent binaries with a giant star and a WD.

In the model MA3, a lower value of $\alpha_{\mathrm{CE}}=0.3$ leads to an increase of the relative fraction of detached LISA NSWD binaries where WDs form first, compared to the cases of $\alpha_{\mathrm{CE}}=1.0$ and $\alpha_{\mathrm{CE}}=3.0$, which can potentially shift the chirp-mass distribution of the whole population of NSWD GW sources to have a higher mass peak. A similar conclusion was drawn by \citet{Korol2023}, who assumed the mass-accretion rate during primordial binary evolution to be limited by the thermal timescale of the accreting secondary, resembling our model MA3.

\subsubsection{The impact of supernova recipes}

\begin{figure*}
	\includegraphics[width=18cm]{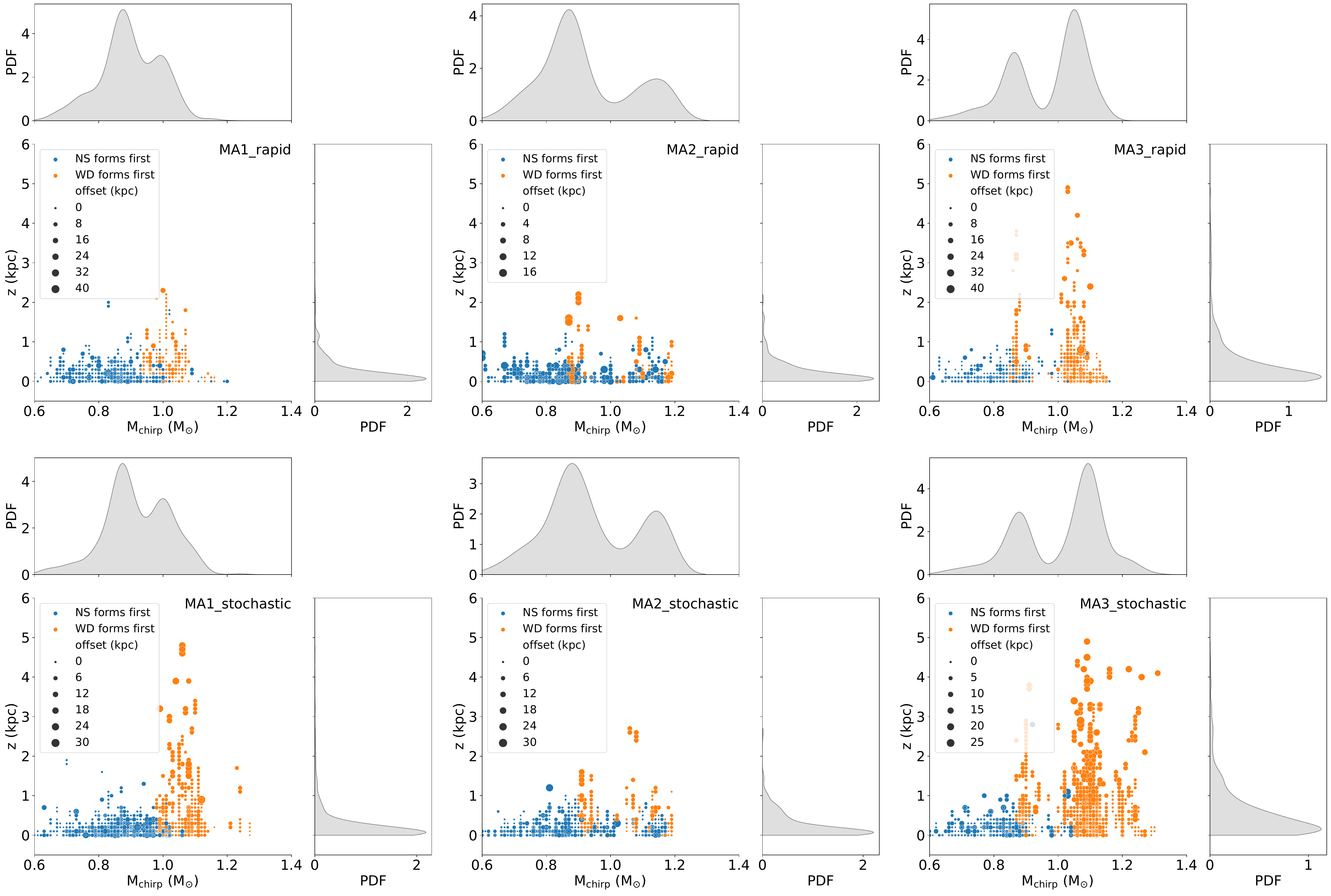}
    \caption{Probability density functions (PDFs) of detached LISA NSWD systems in the Milky Way, as a function of chirp mass and vertical distance from the Galactic plane. The left, middle, and right panels correspond to the models MA1, MA2, and MA3, respectively. The top and bottom panels represent the rapid and stochastic supernova mechanisms, respectively. Here we adopt $\alpha_{\mathrm{CE}}=1.0$. In each panels, the blue dots denote systems where NSs form first, while the orange dots denote systems where WDs form first. The size of each dot reflects its offset distance from the Galactic center.} 
    \label{fig:location}
\end{figure*}

Fig. \ref{fig:location} shows the probability density functions of detached LISA NSWDs in the Milky Way, as a function of chirp mass $M_{\rm chirp}$ and vertical distance $z$ from the Galactic plane. Here, we adopt $\alpha_{\mathrm{CE}}=1.0$. The blue and orange dots with big scatters correspond to the NSWD systems formed via Scenarios 1 and 2, respectively. Different sizes of these dots represent their offset distances from the Galactic center. The left, middle, and right panels represent the models MA1, MA2, and MA3, respectively. The top and bottom panels represent the rapid and stochastic supernova mechanisms, respectively. There is a tendency that the vast majority of the binaries where NSs form first are distributed with $z < 1\,\rm kpc$, while the  systems where WDs form first are more likely to locate at relatively large $z$ with a tail up to $\sim 2-5\,\rm kpc$. And, the latter systems are expected to have significantly large offset distances from the Galactic center. The main reason for these discrepancies is that the systems where WDs form first are more susceptible to the kick velocities of natal NSs. Also, we should note that a significant fraction of LISA NSWD binaries where WDs form first have eccentric orbits.

For the supernova mechanisms changing from the rapid to the stochastic recipes, we observe a slight shift in the overall distribution of chirp masses of LISA NSWD sources towards higher values, especially in the model MA3. This shift is because the stochastic mechanism can produce more massive NSs with masses  of $\sim 1.2 M_{\odot}- 1.6 M_{\odot}$ than the rapid mechanism that forms $\sim 1.1$ NSs. Furthermore, the stochastic mechanism allows part of NSs to have small kick velocities, compared to the rapid mechanism, therefore preventing the disruption of more binaries where WDs form first during supernova explosions. As a result, we can see for the stochastic mechanism that some LISA sources locate at relatively large $z$. However, this difference from the spatial locations of LISA NSWD binaries cannot provide strong constraints on our adopted supernova mechanisms. As pointed out by \citet{Korol2023}, the majority of LISA NSWDs locate within $z=5 \mathrm{~kpc}$ from the Galactic disk. Another avenue of exploration lies in decoupling the component masses of observable binaries, which may offer improved constraints on the remnant masses after supernova explosions.

\subsection{Interacting NSWD Binaries}  \label{sec:interacting}

\begin{figure}
	\includegraphics[width=\columnwidth]{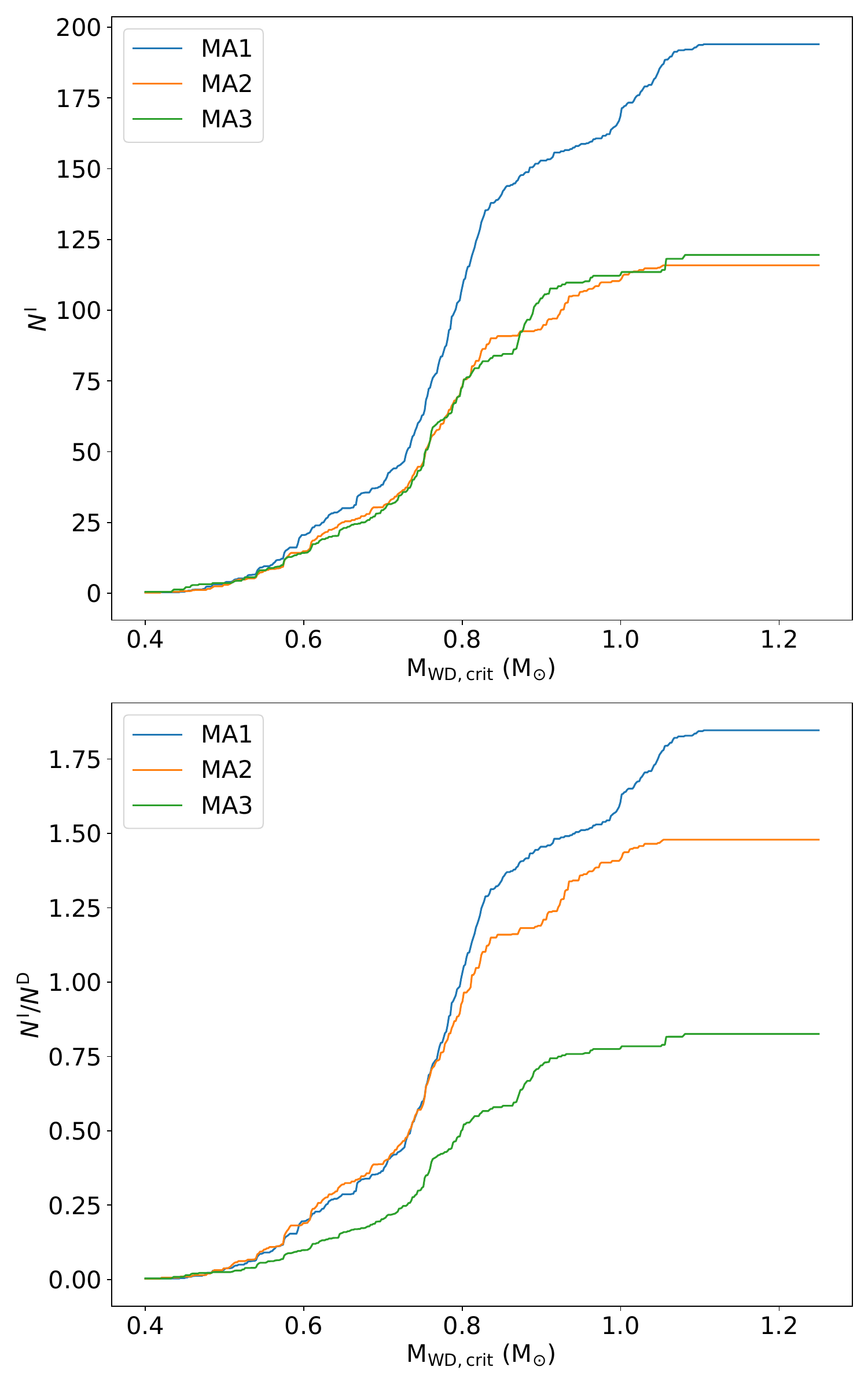}
    \caption{The influence of $M_{\mathrm{WD, crit}}$ on calculated numbers of interacting LISA NSWD binaries (top panel), as well as number ratios of interacting systems to detached systems (bottom panel). The blue, orange, and green curves correspond to the models MA1, MA2, and MA3, respectively. Here $\alpha_{\mathrm{CE}}=1.0$ and the rapid supernova mechanism are adopted.} 
    \label{fig:polyline}
\end{figure}

For close detached NSWDs with orbital periods less than $\sim 0.4$ days, in  particular eccentric systems, the emission of GW is able to significantly shrink their orbits. Within a Hubble time, RLOF occurs in these binaries, leading to the formation of interacting systems. Depending on the stability of mass transfer  via RLOF, NSWD binaries may either merge or evolve into stable UCXBs. Notably, a large value of $M_{\mathrm{WD, crit}}$ can increase the number of interacting NSWD binaries.

Fig. \ref{fig:polyline} shows the expected number of interacting LISA NSWDs, as well as the number ratio of interacting systems to detached systems, as a function of  $M_{\mathrm{WD, crit}}$. In this analysis, we choose $\alpha_{\mathrm{CE}}=1.0$ and the rapid supernova mechanism. Three mass accretion models (i.e. MA1, MA2, and MA3) are taken into account for comparison. Across these models, a pronounced increase in the number of observable sources occurs near $0.8 M_{\odot}$, as this is where the mass distribution of WDs has a peak (see also Fig. \ref{fig:detached}). In each model, there is a plateau at the high-mass end of $\sim 1.1-1.2M_\odot$. This plateau arises because the orbital shrinkage of the systems with $\gtrsim 1.1 M_{\odot}$ ONeWDs caused by GW radiation dominates and these systems always merge, although stable mass transfer is allowed to happen. Overall, when $M_{\mathrm{WD, crit}}$ is below 0.6, the three models exhibit a similar trend. For $M_{\mathrm{WD, crit}}$ exceeding 0.6, the model MA1 is able to produce more interacting NSWD systems, since it allows the formation of more close binaries with less-massive COWDs compared to the models MA2 and MA3. 

In total, the number ratios of interacting to detached NSWD binaries are less than about
1.9/1.5/0.9, corresponding to the models MA1/MA2/MA3, respectively. Importantly, these number ratios are sensitive to the options of $M_{\mathrm{WD, crit}}$. 
In principle, we can provide some constraints on $M_{\mathrm{WD, crit}}$ if a number of interacting and detached LISA NSWD binaries with COWD/ONeWD components are identified in the future. However, it is essential to consider the formation of interacting LISA sources via other pathways such as the RLOF and the AIC channels. Distinguishing between these channels becomes challenging. On the one hand, all detectable interacting NSWDs exhibit eccentricities of $< 0.001$ which are below the measurement threshold of approximately 0.1 by LISA \citep{Korol2023}. On the other hand, the RLOF channel can produce NS$-$HeWD systems with $z>1 \mathrm{~kpc}$, such as J0348+0432 \citep{Antoniadis2013}. It is possible to distinguish the RLOF and the CE channels if one can observe different spectra with components from originally HeWDs or COWDs/ONeWDs. Furthermore, the AIC channel can lead to the formation of both NS$-$HeWD and NS$-$COWD systems (see Section \ref{sec:AIC Mechanism} for more details).

\subsection{Merger rates}
\label{sec:merger}

The merger rates ($R_{\mathrm{merger}}$) of Galactic NSWD binaries under various models are presented in Table \ref{tab:total number}, varying in the range of about $ 5-90\,\rm Myr^{-1}$. It is worth noting that the merger rates are relatively low in the models with $\alpha_{\mathrm{CE}}=0.3$, which are consistent with the small numbers of corresponding LISA binaries predicted in these models. In the models MA1 and MA2, increasing $M_{\mathrm{WD, crit}}$ from $0.2M_\odot$ to $1.25M_\odot$ can greatly decrease the merger rates by a factor of $\sim4-8$. However, the model MA3 shows an exception, as it allows the formation of a substantial number of NS$-$ONeWD binaries, which are bound to merge despite a large value of $M_{\mathrm{WD, crit}}$.

It is believed that NSWD mergers are relevant to some observable transients outside the Milky Way \citep[e.g.,][]{Bobrick2022,Kaltenborn2023}. One can calculate the merger rate density ($\mathcal{R}_{\mathrm{merger}}$) of NSWD binaries in the local Universe, if simultaneously modeling the evolution of star formation and metallicity as a function of redshift. Here we present a rough estimation of  $\mathcal{R}_{\mathrm{merger}}$, according to our obtained $R_{\mathrm{merger}}$. We assume that the number density of Milky Way equivalent galaxies within the local Universe is $0.01\rm\,Mpc^{-3}$ \citep[e.g.,][]{Abadie2010}. This produces a conversion factor between $R_{\mathrm{merger}}=1\rm\, Myr^{-1}$ and $\mathcal{R}_{\mathrm{merger}}=10\rm\, Gpc^{-3}\,yr^{-1}$. So we estimate the merger rate density of NSWD binaries in the local Universe of $ \sim 50-900\rm\, Gpc^{-3}\,yr^{-1}$, which is agreement with the theoretical prediction of $ 390\rm\,Gpc^{-3}\,yr^{-1}$ made by \citet{Zhao2021}.

\section{Discussion} \label{sec:discussion}

\subsection{Identification of NSWDs}

Although the detection of Galactic NSWD binaries as GW sources can provide valuable information to constrain the origin of these sources and the physics of binary interaction, not all NSWDs can be discerned among numerous GW sources in the Milky Way.

The identification of LISA NSWD systems typically relies on the measurement of chirp masses, which usually range from approximately $0.35$ to $1.2 M_{\odot}$ \citep[e.g.,][]{Korol2023}. For comparison, the chirp mass distribution of WDWD and NSNS systems exhibits a peak around $\sim 0.25-0.4 \mathrm{M}_{\odot}$ \citep{Korol2022} and $1.1 M_{\odot}$ \citep{Korol2021}, respectively. The determination of chirp masses can often be inferred through the measurement of GW frequency derivative $\dot{f}_{\mathrm{GW}}$ for nearby GW sources with exceptionally high S/N. Additionally, for other systems, \citet{Tauris2018} indicated that combining measurements of optical distance and GW strain amplitude can effectively constrain $\dot{f}_{\mathrm{GW}}$. Furthermore, eccentricities can serve as an alternative feature to distinguish NSWD systems from WDWD systems in cases where chirp masses cannot be directly measured. \citet{Korol2023} demonstrated that the minimum detectable eccentricity, derived from full Bayesian parameter estimation \citep{Moore2023}, can reach as low as $\sim 0.03$ when searching for realistic eccentric NSWD systems. Moreover, as mentioned by \citet{Tauris2018}, a direct method to identify NSWD systems is the search of both optical WDs and radio pulsars.

\subsection{The degeneracy of multiple parameters} \label{sec:influences factors}

In our study, we include multiple parameters related to binary evolution, which can potentially cross-influence the population statistics of LISA NSWD systems and lead to possible degeneracy of these parameters.

The mass accretion models (MA1, MA2 and MA3) play a vital role in determining the formation order of NS/WD, thereby affecting the eccentricity distribution of detached LISA NSWD binaries. The formation of the systems with eccentricities larger than 0.1 is sensitive to the option of these models. Additionally, the mass accretion models can also influence the chirp mass distributions, as the NSWD binaries where WDs from first tend to contain more massive WDs than the systems where NSs form first. Combining the distributions of orbital eccentricities and chirp masses for detached LISA NSWD binaries is able to provide constraints on the mass accretion models.

Varying CE ejection efficiencies $\alpha_{\mathrm{CE}}$ can lead to different relative fractions of detached LISA NSWD binaries where NSs or WDs form first. The impact of $\alpha_{\mathrm{CE}}$ on the ratios of $N^{\rm D}_{\rm e>0.1}$ to $N^{\rm D} $ is considerably smaller than that of mass accretion models (see Fig. \ref{fig:pie}). Furthermore, disentangling the impact of $\alpha_{\mathrm{CE}}$ from that of supernova mechanisms is challenging, as both of them can significantly influence the calculated numbers and the obtained parameter distributions of detached LISA NSWD binaries (see Fig. \ref{fig:detached} and Figs. \ref{fig:detached_CE03}$-$\ref{fig:detached_SN} in Appendix A).

\subsection{AIC Mechanism} \label{sec:AIC Mechanism}

\begin{figure}
	\includegraphics[width=\columnwidth]{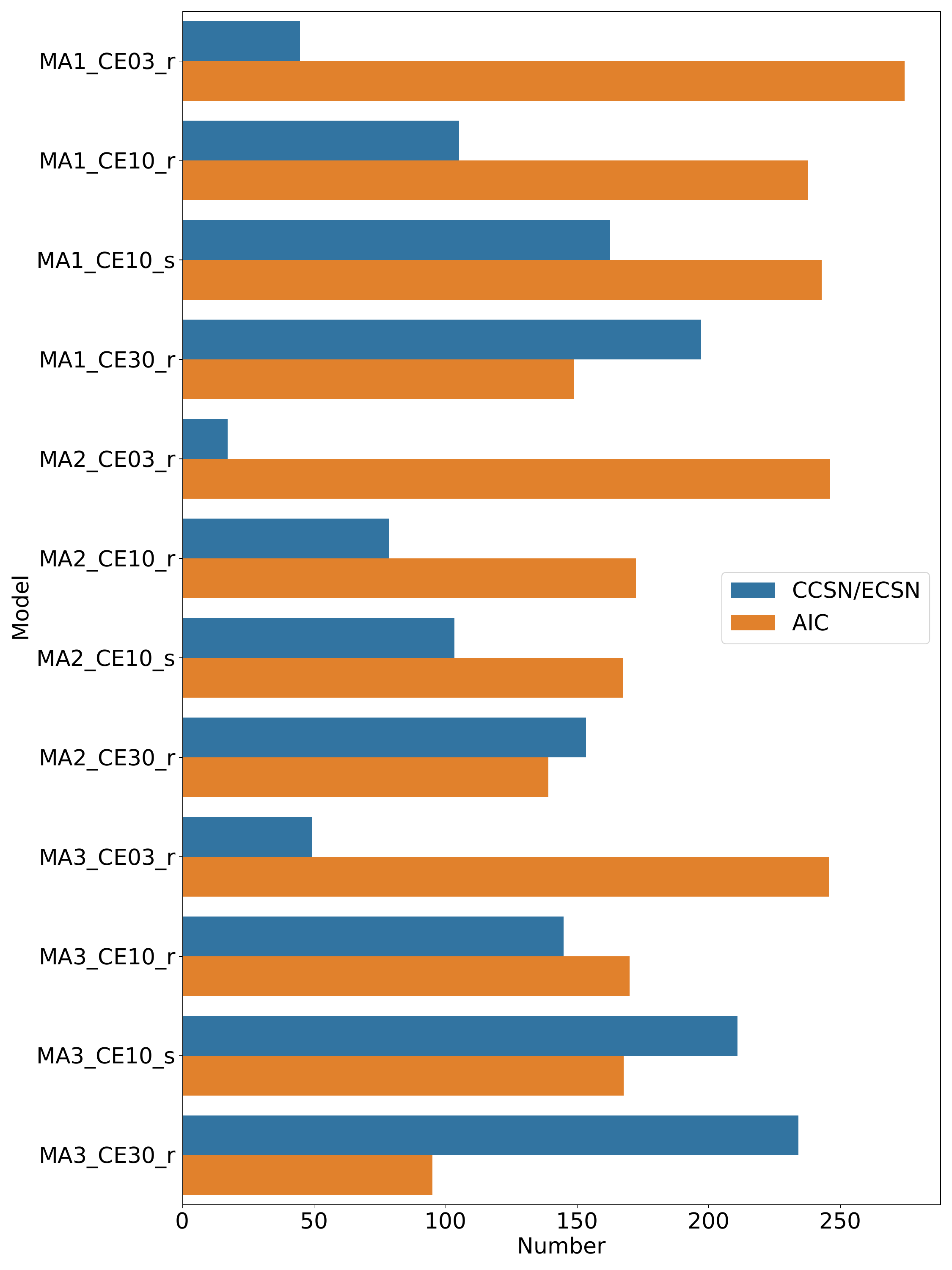}
    \caption{The bar chart illustrates calculated numbers of LISA NSWD binaries under various models. The blue bars represent the systems with  CCSN/ECSN NSs, while the orange bars represent the systems with AIC NSs. Since the criterion of $M_{\mathrm{WD, crit}}=0.2M_\odot$ is used, all LISA binaries with CCSN/ECSN NSs from our calculations are detached systems. Here, the  LISA binaries with AIC NSs include interacting systems.} 
    \label{fig:bar}
\end{figure}

Based on our calculations, we observe a significant contribution from the NSWD binaries with AIC NSs to the whole population of Galactic LISA NSWD sources. This contribution was not considered in previous analyses due to considerable uncertainties of the AIC mechanism itself. 

Beginning the evolution from a primordial binary, the primary star firstly leaves an ONeWD. When the secondary star evolves to the giant branch and fills its Roche lobe, a CE phase is triggered. This phase leads to the formation of the ONeWD binary with a helium star or another WD companion. Subsequently, the ONeWD collapse into an NS when its mass exceeds the critical mass of $ 1.38M_{\odot}$ due to accretion. 

In Fig. \ref{fig:bar}, we provide an estimate of the number of LISA NSWD binaries with AIC NSs in the Milky Way. In contrast to the systems with CCSN/ECSN NSs, a lower value of $\alpha_{\mathrm{CE}}$ tends to produce more LISA sources with AIC NSs. 
Overall, our models predict that the Milky Way may host about $100-300$ LISA NSWD binaries with AIC NSs.

The AIC channel is not yet fully understood, including uncertainties from the treatments of mass-transfer stability in binaries with WD accretors and mass-retention efficiency of accreting WDs. For the stability of mass transfer between giant-like donor stars and WDs, we adopt the default critical mass ratio in the BSE code, as Equation (56) of \citet{Hurley2002}. This criterion of mass-transfer stability is derived by \citet{Hjellming1987}, assuming the transfer of mass and orbital angular momentum is conservative during the evolution. However, the realistic value of $\zeta_{\mathrm{RL}}$ depends on specific mass-loss mechanisms when mass transfer is non-conservative \citep{Soberman1997}. The default criterion involving a low critical mass ratio implies that almost all of the systems with an ONeWD and a giant donor experience a CE phase and evolve to be close binaries, thereby significantly contributing to the population of LISA NSWD binaries during subsequent evolution. However, \citet{Pavlovskii2015} proposed that the critical mass ratio for stable mass transfer in binaries with a giant donor and a compact object ranges from 1.5 to 2.2. More recently, \citet{Ge2020} showed that mass transfer in binaries with giant donor stars is more stable than previously believed.

Furthermore, for stable mass transfer between two WDs, we adopt the critical mass ratio of 0.628 \citep{Hurley2002}, which is similar to the value proposed by \citet{Nelemans2001a}. This critical mass ratio allows most ONeWDs to effectively accrete mass from HeWDs or COWDs, eventually resulting in the formation of NSs via AIC. Additionally, we assume that steady nuclear burning occurs on the surface of accreting WDs if mass-transfer rate falls within the range of $(1.03 - 2.71) \times 10^{-7} \mathrm{M}_{\odot} \mathrm{yr}^{-1}$. However, this range greatly depends on the masses and the temperatures of the WDs involved, as well as the components of accreted material. Therefore, there remains uncertainty regarding whether the AIC mechanism significantly contributes the population of LISA NSWD binaries.

\section{Conclusions} \label{sec:conclusion}

In this study, we have performed binary population synthesis calculations to investigate the origins of LISA NSWD binaries in the Milky Way and examine the influences of different assumptions related to binary evolution on their characteristic distribution. Our results reveal that approximately 17$-$234 detached NSWD binaries and less than 200 interacting systems can serve as detectable LISA sources, excluding the NSWDs originating from the RLOF and the AIC channels.

The model MA3, with near-conservative mass transfer during primordial binary evolution, predicts most of detached LISA NSWD binaries are systems where WDs form first (see Fig. \ref{fig:pie}).  Among all detached LISA binaries from our calculations, the fraction for the systems with eccentricities larger than 0.1 can reach as high as $\sim 0.3-0.4$.
While for the models MA1 and MA2, we obtain that $\lesssim 0.2$ of detached LISA NSWD systems have eccentricities of larger than 0.1. These eccentric systems are more likely to locate at large vertical
distances from the Galactic plane. Furthermore, detached LISA NSWD binaries are more likely to host a massive ONeWD in the model MA3, compared to the models MA1 and MA2. By studying the distributions of the binary parameters and the spatial locations of detached LISA NSWD sources, it is possible to constrain mass accretion models (i.e., MA1, MA2, and MA3). 

Also, we have demonstrated that CE ejection efficiencies  $\alpha_{\mathrm{CE}}$ can significantly influence the expected numbers of LISA NSWD sources in the Milky Way. A smaller value of $\alpha_{\mathrm{CE}}$ tends to suppress the formation of the systems with eccentricities of $>0.1$, particularly in the models MA1 and MA2. 
For supernova mechanisms, the adoption of the stochastic recipe tends to generate more binaries with large chirp masses, in contrast to the rapid recipe.  

At last, we propose that the stability of mass transfer between WDs and NSs could be constrained, according to the observations of interacting LISA NSWD binaries also appearing as UCXBs.

\section*{Acknowledgements}

We thank the anonymous referee for helpful suggestions that improved this paper. This work was supported by the National Key Research and Development Program of China (Grant Nos. 2023YFA1607902, 2021YFA0718500), the Natural Science Foundation of China (Nos. 12041301, 12121003, 12373034), the Strategic Priority Research Program of the Chinese Academy of Sciences (Grant No. XDB0550300), and the Project U1838201 supported by NSFC and CAS.

%%%%%%%%%%%%%%%%%%%%%%%%%%%%%%%%%%%%%%%%%%%%%%%%%%
\section*{Data Availability}

The data underlying this article will be shared on reasonable request to the corresponding author.

%%%%%%%%%%%%%%%%%%%% REFERENCES %%%%%%%%%%%%%%%%%%

% The best way to enter references is to use BibTeX:

\bibliographystyle{mnras}
\bibliography{article} % if your bibtex file is called example.bib

% Alternatively you could enter them by hand, like this:
% This method is tedious and prone to error if you have lots of references
%\begin{thebibliography}{99}
%\bibitem[\protect\citeauthoryear{Author}{2012}]{Author2012}
%Author A.~N., 2013, Journal of Improbable Astronomy, 1, 1
%\bibitem[\protect\citeauthoryear{Others}{2013}]{Others2013}
%Others S., 2012, Journal of Interesting Stuff, 17, 198
%\end{thebibliography}

%%%%%%%%%%%%%%%%%%%%%%%%%%%%%%%%%%%%%%%%%%%%%%%%%%

%%%%%%%%%%%%%%%%% APPENDICES %%%%%%%%%%%%%%%%%%%%%

\appendix

\section{Parameter distributions of detached LISA NSWD binaries}

\begin{figure*}
	\includegraphics[width=18cm]{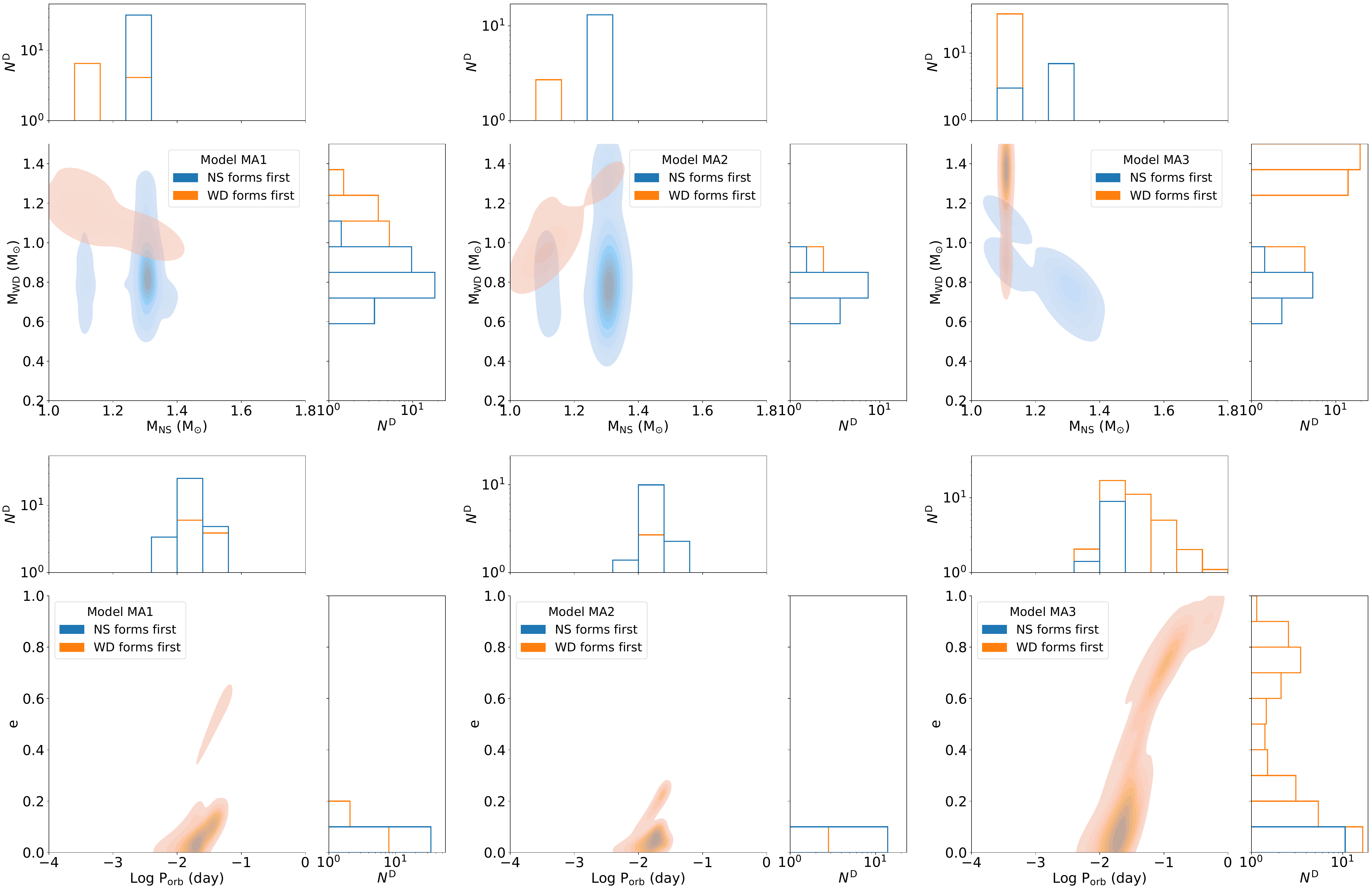}
    \caption{Similar to Fig. \ref{fig:detached}, but assuming $\alpha_{\mathrm{CE}} = 0.3$.} 
    \label{fig:detached_CE03}
\end{figure*}

\begin{figure*}
	\includegraphics[width=18cm]{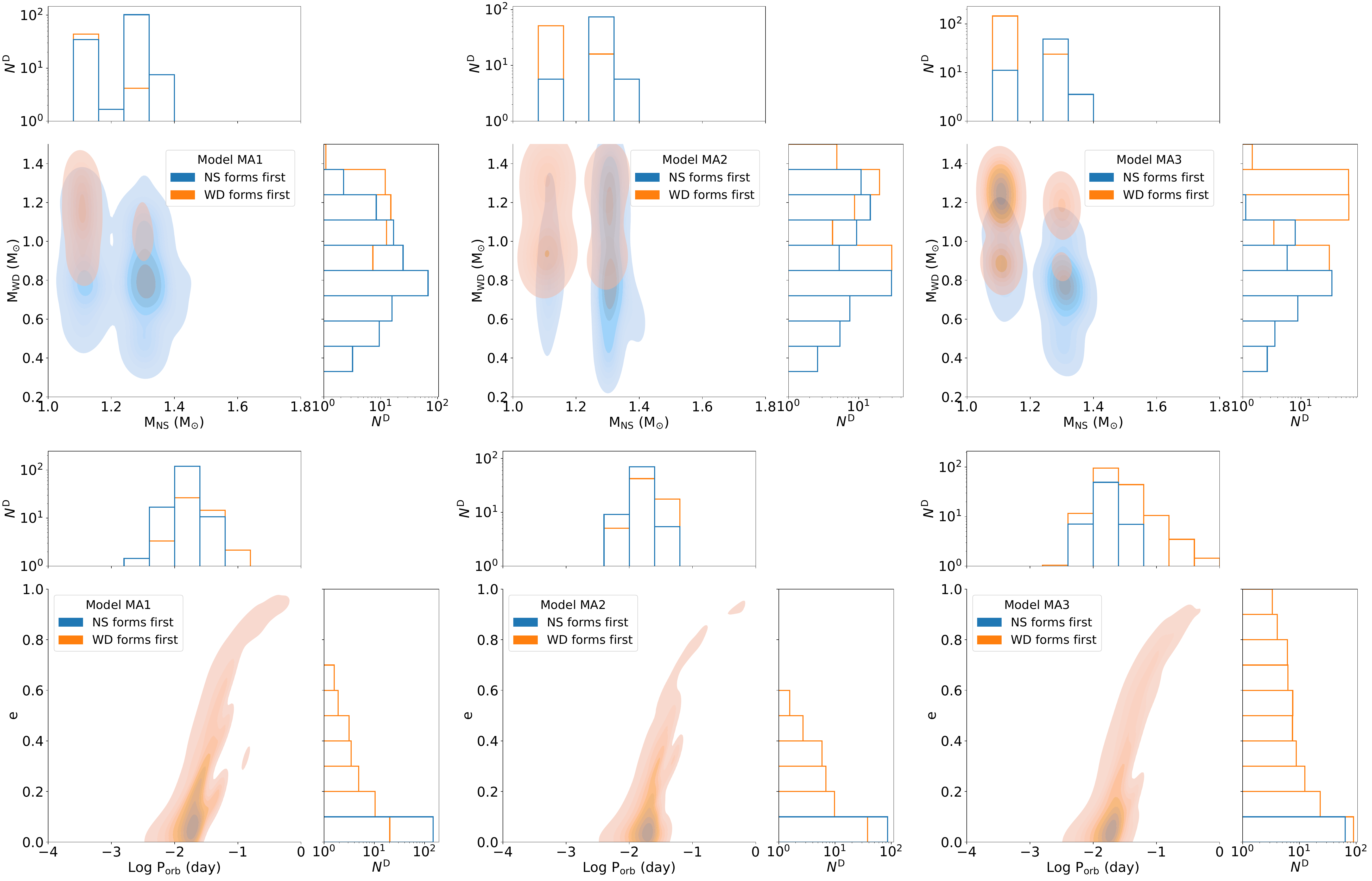}
    \caption{Similar to Fig. \ref{fig:detached}, but assuming $\alpha_{\mathrm{CE}} = 3.0$.} 
    \label{fig:detached_CE30}
\end{figure*}

\begin{figure*}
	\includegraphics[width=18cm]{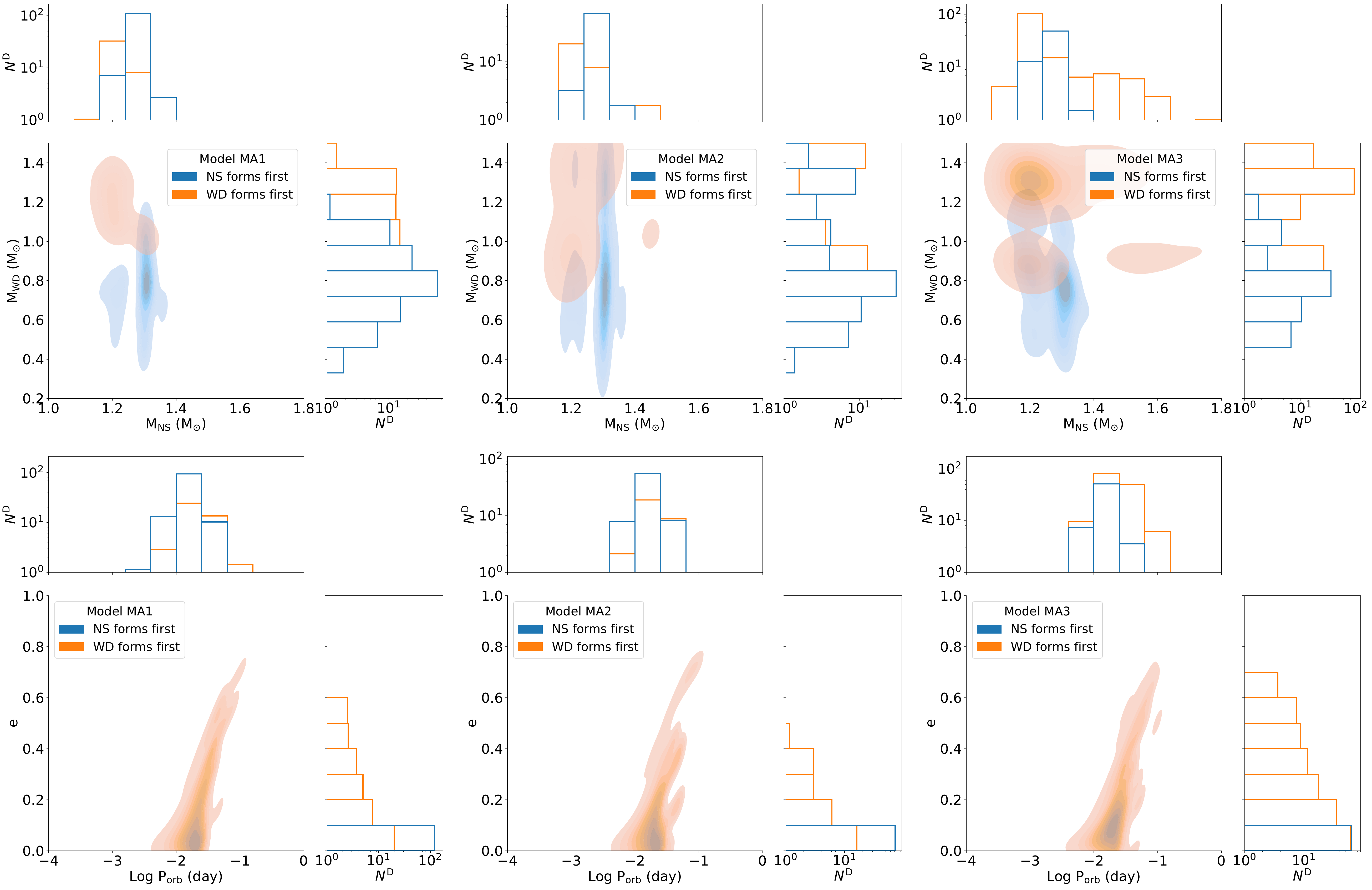}
    \caption{Similar to Fig. \ref{fig:detached}, but assuming the stochastic mechanism of supernova explosions.} 
    \label{fig:detached_SN}
\end{figure*}

Figs. \ref{fig:detached_CE03}$-$\ref{fig:detached_SN} show calculated number distributions of Galactic LISA sources of detached NSWD systems as a function of binary parameters, similar to Fig. \ref{fig:detached}, by assuming different CE efficiencies and supernova recipes.

% If you want to present additional material which would interrupt the flow of the main paper,
% it can be placed in an Appendix which appears after the list of references.

%%%%%%%%%%%%%%%%%%%%%%%%%%%%%%%%%%%%%%%%%%%%%%%%%%

% Don't change these lines
\bsp	% typesetting comment
\label{lastpage}
\end{document}